\documentclass{jpsj3}
\usepackage{graphicx}
\usepackage{dcolumn}
\usepackage{bm}
\usepackage{color}

\title{%
Quasiclassical Theory on Third-Harmonic Generation 
in Conventional Superconductors 
with Paramagnetic Impurities
}
\author{%
Takanobu Jujo
\thanks{E-mail address: jujo@ms.aist-nara.ac.jp}
}
\inst{%
Graduate School of Materials Science, 
Nara Institute of Science and Technology, 
Ikoma, Nara 630-0101, Japan
}
\recdate{September 27, 2017}
\abst{%
We investigate the third-harmonic generation (THG) of 
$s$-wave superconductors under microwave pulse irradiation. 
We consider the effect of paramagnetic impurities on the THG 
intensity of dirty superconductors. 
The nonlinear response function is calculated using the method 
of the quasiclassical Green function. It is shown that 
the amplitude mode is included as the vertex correction and 
makes a predominant contribution 
to the THG intensity. 
When the effect of paramagnetic impurities is weak, 
the THG intensity shows a peak at the temperature at which 
the superconducting gap is about the same as the frequency of 
the incident pulse, similarly to in experiments. 
As the effect of paramagnetic impurities is strengthened, 
the peak of the THG intensity disappears. 
This indicates that time-reversal symmetry breaking due to 
paramagnetic impurities eliminates the well-defined amplitude mode. 
The result of our calculation shows that the existence of 
the amplitude mode can be confirmed through the THG intensity. 
The result of a semiquantitative calculation is in good 
agreement with the experimental result, and it also shows 
that the diamagnetic term is negligible. 
}

\begin{document}
\setlength{\textwidth}{504pt}
\setlength{\columnsep}{14pt}
\hoffset-23.5pt
\maketitle

\section{Introduction}

In recent years, studies on the nonlinear optical 
response in superconductors have advanced 
with the development of microwave spectroscopy. 
Many of these studies were on the dynamics 
of transient responses using pump-probe spectroscopy, 
and the main objective was to elucidate the interaction 
effect in superconductors including 
strongly correlated electron systems. 
(For a review, see Ref. 1
%~\cite{giannetti} 
for example.) 

Another important aspect of the nonlinear optical response 
in the research of superconductors is that we can 
investigate phenomena that do not appear in the linear response. 
The amplitude fluctuation of the superconducting 
order parameter~\cite{schmid,volkov} 
is a typical example because the amplitude mode 
is not reflected in the linear absorption spectrum. 
(This is different from the phenomenon that 
the phase mode is pushed up to a high energy 
by the long-range Coulomb interaction.~\cite{anderson58}) 

So far, the amplitude mode has been observed in the phonon spectrum 
through the coupling of the superconductivity and 
the charge density wave in a superconductor 
coexisting with the charge density wave state.~\cite{sk,littlewood} 
Recently, it has been reported that 
the amplitude mode can be observed in ordinary superconductors 
by pump-probe spectroscopy~\cite{matsunaga} and 
third-harmonic generation (THG).~\cite{matsunaga14} 
The former experimental study asserted that the observed 
transient oscillation 
of optical conductivity is an expression of the amplitude mode. 
The latter showed that the peak appears at 
a temperature at which the superconducting gap is 
equal to the frequency of the incident wave. 
This may indicate the existence of the amplitude mode.~\cite{note0} 

Theoretically, the excitation of the amplitude mode 
overlaps the quasiparticle excitation 
by two-photon absorption in the spectrum.~\cite{jujo15} 
Thus, it is required to distinguish these two phenomena 
in order to confirm the amplitude mode. 
As a proposal for identifying the amplitude mode, 
in this paper we investigate the THG intensity 
in superconductors with paramagnetic impurities. 
Time-reversal symmetry breaking due to paramagnetic impurities 
leads to instability of the amplitude mode.~\cite{ivlev} 
Therefore, the peak of the THG intensity will decrease 
when the amplitude mode is dominant in the THG intensity. 
In addition, in order to clarify the existence of 
the amplitude mode, 
we conduct a semiquantitative evaluation, 
which has not been done so far 
(for example, supplementary materials of 
Refs. 8 and 12 
%~\cite{matsunaga14,matsunaga17} 
and references therein), 
and compare it with the experimental result.

The following is the structure of this paper. 
Section 2 gives the formulation by the quasi-classical 
approximation for calculating the response function. 
Section 3 gives an expression for the THG intensity 
under pulse irradiation and shows 
the results of a numerical calculation such as 
its temperature dependence. 
A semiquantitative evaluation is also given. 
Section 4 gives an expression for the diamagnetic term 
and shows that this term is negligible in 
the THG intensity.

\section{Formulation}

A nonlinear current is written using the 
quasiclassical Green function ($g^K$) as follows:~\cite{eliashberg} 
\begin{equation}
J^{\mu}_{\omega}=e\int_{\rm FS}v^{\mu}_k
\frac{mk_F}{2\pi}\int\frac{d\epsilon}{4\pi i}
\sum_{\nu=x,y,z}
{\rm Tr}[v_k^{\nu}\hat{g}^{K(3)}_{\nu}(\epsilon+\omega,\epsilon)]. 
\end{equation}
Here, $v_k$ and $m$ are 
the velocity and mass of electrons, respectively, and 
$k_F$ is the Fermi wave number. (We put $\hbar=c=1$ in this paper with 
$c$ the velocity of light.) 
$\int_{\rm FS}$ means integration on the Fermi surface. 
The superscripts $K$ and $(3)$ in $\hat{g}^{K(3)}$ indicate 
the Keldysh Green function~\cite{keldysh} 
and the third order of the external fields, 
respectively. The modifier `` $\hat{}$ ''
means a matrix in Nambu representation. 
$\mu,\nu=x,y,z$ indicate spatial directions. 
We assume the isotropic case and omit this index 
hereafter. 

$\hat{g}^{K(3)}(\epsilon+\omega,\epsilon)$ 
is derived from kinetic equations in the dirty limit.~\cite{usadel,kopnin} 
In this limit 
the quasiclassical Green function can be divided into 
an odd order and an even order with respect to the external field. 
In the kinetic equation for the former, 
the effect of scattering by nonmagnetic 
impurities does not vanish in the collision integral and 
is predominant over other terms such 
as the superconducting gap. 
Thus, the kinetic equation in this case 
can be solved, and its solution 
is written as 
\begin{equation}
\begin{split}
&\hat{g}^{K(3)}(\epsilon+\omega,\epsilon)
=e\int\frac{d\omega'}{2\pi}
A_{\omega'}
\frac{-1}{2\alpha}
\bigl(
\hat{g}^+_{\epsilon+\omega}\hat{g}^{K}_{\epsilon+\omega-\omega',\epsilon}
+
\hat{g}^K_{\epsilon+\omega}\hat{g}^{-}_{\epsilon+\omega-\omega',\epsilon}
+
\hat{g}^{+}_{\epsilon,\epsilon-\omega+\omega'}\hat{g}^K_{\epsilon-\omega}
+
\hat{g}^{K}_{\epsilon,\epsilon-\omega+\omega'}\hat{g}^-_{\epsilon-\omega} \\
&-
\hat{\tau}_3\hat{g}^+_{\epsilon}
\hat{\tau}_3\hat{g}^{K}_{\epsilon,\epsilon-\omega+\omega'}
-
\hat{g}^K_{\epsilon}\hat{\tau}_3
\hat{g}^{-}_{\epsilon,\epsilon-\omega+\omega'}\hat{\tau}_3
-
\hat{\tau}_3\hat{g}^{+}_{\epsilon,\epsilon-\omega+\omega'}
\hat{\tau}_3\hat{g}^K_{\epsilon-\omega+\omega'} 
-
\hat{g}^{K}_{\epsilon,\epsilon-\omega+\omega'}\hat{\tau}_3
\hat{g}^-_{\epsilon-\omega+\omega'}\hat{\tau}_3
\bigr). 
\end{split}
\end{equation}
Here, $\alpha=(mk_F/2\pi)n_i u_i^2$ with 
$n_i$ the concentration of nonmagnetic impurities and 
$u_i$ the magnitude of the potential. 
The dirty limit means that $\alpha\gg \Delta$. 
$A_{\omega}$ is the external vector potential. 
$\hat{\tau}_3=\left(
\begin{smallmatrix}1 & 0\\ 0 & -1\end{smallmatrix}\right)$. 
$\hat{g}_{\epsilon}$ is the quasiclassical Green function 
in the equilibrium state and 
$\hat{g}_{\epsilon,\epsilon'}$ is a second-order function on external 
fields ($+$ and $-$ in the superscript indicate 
the retarded and advanced Green function, respectively). 
The kinetic equations for the latter are written as 
\begin{equation}
\begin{split}
&\hat{\tau}_3\epsilon\hat{g}^+_{\epsilon,\epsilon'}
-\hat{g}^+_{\epsilon,\epsilon'}\epsilon'\hat{\tau}_3
-
\left[
\hat{\tau}_3
\hat{\Sigma}^+_{\epsilon}
\hat{g}^+_{\epsilon,\epsilon'}
-\hat{g}^+_{\epsilon,\epsilon'}
\hat{\Sigma}^+_{\epsilon'}
\hat{\tau}_3
+
\hat{\tau}_3
\hat{\Sigma}^+_{\epsilon,\epsilon'}
\hat{g}^+_{\epsilon'}
-\hat{g}^+_{\epsilon}
\hat{\Sigma}^+_{\epsilon,\epsilon'}
\hat{\tau}_3
\right] 
\\
&-e^2D_{\alpha}
\int\frac{d\omega_1d\omega_2}{(2\pi)^2}
A_{\omega_1}A_{\omega_2}
\left[
\hat{\tau}_3\hat{g}^+_{\epsilon-\omega_1}
\hat{g}^+_{\epsilon'}
-\hat{g}^+_{\epsilon}
\hat{g}^+_{\epsilon'+\omega_1}
\hat{\tau}_3
\right]\delta(\epsilon-\epsilon'-\omega_1-\omega_2)
=0 
\label{eq:kefor+}
\end{split}
\end{equation}
and 
\begin{equation}
\begin{split}
&\hat{\tau}_3\epsilon\hat{g}^{(a)}_{\epsilon,\epsilon'}
-\hat{g}^{(a)}_{\epsilon,\epsilon'}\epsilon'\hat{\tau}_3
-
\left[
\hat{\tau}_3
\hat{\Sigma}^+_{\epsilon}
\hat{g}^{(a)}_{\epsilon,\epsilon'}
-\hat{g}^{(a)}_{\epsilon,\epsilon'}
\hat{\Sigma}^-_{\epsilon'}
\hat{\tau}_3
+
\hat{\tau}_3
\hat{\Sigma}^{(a)}_{\epsilon,\epsilon'}
\hat{g}^-_{\epsilon'}
-\hat{g}^+_{\epsilon}
\hat{\Sigma}^{(a)}_{\epsilon,\epsilon'}
\hat{\tau}_3
\right] \\
&-e^2D_{\alpha}
\int\frac{d\omega_1d\omega_2}{(2\pi)^2}
A_{\omega_1}A_{\omega_2}
\Bigl\{
\hat{\tau}_3
\left[
(t^h_{\epsilon-\omega_1}-t^h_{\epsilon-\omega_1-\omega_2})
\hat{g}^+_{\epsilon-\omega_1}
+
(t^h_{\epsilon}-t^h_{\epsilon-\omega_1})
\hat{g}^-_{\epsilon-\omega_1}
\right]
\hat{g}^-_{\epsilon'} \\
&-
\hat{g}^+_{\epsilon}
\left[
(t^h_{\epsilon'+\omega_1}-t^h_{\epsilon'})
\hat{g}^+_{\epsilon'+\omega_1}
+
(t^h_{\epsilon'+\omega_1}-t^h_{\epsilon})
\hat{g}^-_{\epsilon'+\omega_1}
\right]
\hat{\tau}_3
\Bigr\}\delta(\epsilon-\epsilon'-\omega_1-\omega_2)
=0. 
\label{eq:kefora}
\end{split}
\end{equation}
Here, $t^h_{\epsilon}:={\rm tanh}(\epsilon/2T)$ ($T$ is temperature), 
$\delta(\cdot)$ is the delta function, and 
$\hat{g}^{(a)}_{\epsilon,\epsilon'}:=
\hat{g}^K_{\epsilon,\epsilon'}
-t^h_{\epsilon'}\hat{g}^+_{\epsilon,\epsilon'}
+t^h_{\epsilon}\hat{g}^-_{\epsilon,\epsilon'}$ 
($\hat{g}^{(a)}=\hat{\Sigma}^{(a)}=\hat{0}$ 
in the equilibrium state).~\cite{eliashberg} 
The effect of impurity scatterings is calculated 
with the Born approximation,~\cite{abrikosov59} 
and the interaction between electrons and phonons 
is treated with the weak-coupling approximation. 
The self-energy is written as 
$\hat{\Sigma}^s_{\epsilon,\epsilon'}=
\hat{\Sigma}^{(ep)s}_{\epsilon,\epsilon'}
+\hat{\Sigma}^{(ni)s}_{\epsilon,\epsilon'}
+\hat{\Sigma}^{(pi)s}_{\epsilon,\epsilon'}$ 
with 
$\hat{\Sigma}^{(ni)s}_{\epsilon,\epsilon'}=
\alpha \hat{\tau}_3\hat{g}^s_{\epsilon,\epsilon'}\hat{\tau}_3$ 
(the effect of nonmagnetic impurity scattering) and 
$\hat{\Sigma}^{(pi)s}_{\epsilon,\epsilon'}=
\alpha_p \hat{g}^s_{\epsilon,\epsilon'}$
(the effect of paramagnetic impurity scattering~\cite{abrikosov61,skalski}) 
[$s=+$, $-$ or $(a)$]. 
$\hat{\Sigma}^{(ep)+}_{\epsilon,\epsilon'}
=\hat{\Sigma}^{(ep)-}_{\epsilon,\epsilon'}
=g_0\int\frac{d\epsilon_1}{2\pi i}
\hat{\tau}_3\hat{g}^K_{\epsilon_1+(\epsilon-\epsilon')/2,
\epsilon_1-(\epsilon-\epsilon')/2}\hat{\tau}_3$ 
(the electron$-$phonon interaction) 
and 
$\hat{\Sigma}^{(ep)(a)}_{\epsilon,\epsilon'}=
(t^h_{\epsilon}-t^h_{\epsilon'})
\hat{\Sigma}^{(ep)+}_{\epsilon,\epsilon'}$. 
$\alpha_p=(mk_F/2\pi)n'_i u'^2$ 
and $g_0=(mk_F/2\pi)(g_{ph}^2/\omega_D)$ 
with $\omega_D$ the Debye frequency and 
$g_{ph}$ the coupling constant between 
electrons and phonons. 
$D_{\alpha}=v_F^2/6\alpha=v_F^2\tau/3$ 
($\tau=1/2\alpha$ is the relaxation time) is 
the diffusion constant. 

Equations (\ref{eq:kefor+}) and (\ref{eq:kefora}) 
are solved by introducing 
\begin{equation}
\hat{g}^s_{\epsilon,\epsilon'}=
g^s_{\epsilon,\epsilon'}\hat{\tau}_0
+f^s_{\epsilon,\epsilon'}\hat{\tau}_1
\label{eq:solg}
\end{equation}
and 
\begin{equation}
\hat{\Sigma}^{s}_{\epsilon,\epsilon'}=
\Sigma^{n,s}_{\epsilon,\epsilon'}\hat{\tau}_0
+\Sigma^{a,s}_{\epsilon,\epsilon'}\hat{\tau}_1. 
\label{eq:sols}
\end{equation}
Here, 
$n$ and $a$ indicate the normal and anomalous 
self-energy, respectively. 
$\hat{\tau}_0=\left(
\begin{smallmatrix}1 & 0\\0 & 1\end{smallmatrix}
\right)$ and 
$\hat{\tau}_1=\left(
\begin{smallmatrix}0 & 1\\1 & 0\end{smallmatrix}
\right)$. 

By solving the above equations using 
$\Sigma^{(ep)+}_{\epsilon}=\Sigma^{(ep)-}_{\epsilon}
=\Delta\hat{\tau}_1$ 
($\Delta$ is the superconducting gap 
with the effect of impurity scattering included), 
the solution is written as 
\begin{equation}
\begin{pmatrix}
g^s_{\epsilon,\epsilon'} \\ f^s_{\epsilon,\epsilon'} 
\end{pmatrix}
:=
e^2D_{\alpha}
\int\frac{d\omega_1d\omega_2}{(2\pi)^2}
A_{\omega_1}A_{\omega_2}\delta(\epsilon-\epsilon'-\omega_1-\omega_2)
\begin{pmatrix}
\bar{g}^s_{\epsilon,\epsilon'(\omega_1,\omega_2)} 
\\ \bar{f}^s_{\epsilon,\epsilon'(\omega_1,\omega_2)} 
\end{pmatrix}. 
\label{eq:GexclA}
\end{equation}
Here, $s=+$, $-$, or $(a)$, and 
\begin{equation}
\begin{pmatrix}
\bar{g}^{\pm}_{\epsilon,\epsilon'(\omega_1,\omega_2)}  \\
\bar{f}^{\pm}_{\epsilon,\epsilon'(\omega_1,\omega_2)} 
\end{pmatrix}
=
\hat{M}^{\pm\pm}_{\epsilon,\epsilon'}
\frac{1}{2}
\left[
\begin{pmatrix}
g^{\pm}_{\epsilon-\omega_1} \\
f^{\pm}_{\epsilon-\omega_1}
\end{pmatrix}
+
\begin{pmatrix}
g^{\pm}_{\epsilon-\omega_2} \\
f^{\pm}_{\epsilon-\omega_2}
\end{pmatrix}
\right]
+
\hat{M}^{\pm\pm}_{\epsilon,\epsilon'}
\begin{pmatrix}
0 \\
\bar{\Sigma}^{a,+}_{\epsilon,\epsilon'(\omega_1,\omega_2)} 
\end{pmatrix} 
\label{eq:gbar+}
\end{equation}
(the double signs correspond), 
and 
\begin{equation}
\begin{split}
\begin{pmatrix}
\bar{g}^{(a)}_{\epsilon,\epsilon'(\omega_1,\omega_2)}  \\
\bar{f}^{(a)}_{\epsilon,\epsilon'(\omega_1,\omega_2)} 
\end{pmatrix}
=&
\hat{M}^{+-}_{\epsilon,\epsilon'}
\frac{1}{2}
\sum_{\omega'=\omega_1,\omega_2}
\left[
(t^h_{\epsilon-\omega'}-t^h_{\epsilon'})
\begin{pmatrix}
g^+_{\epsilon-\omega'} \\
f^+_{\epsilon-\omega'}
\end{pmatrix}
+
(t^h_{\epsilon}-t^h_{\epsilon-\omega'})
\begin{pmatrix}
g^-_{\epsilon-\omega'} \\
f^-_{\epsilon-\omega'}
\end{pmatrix}
\right]  \\
&+
\hat{M}^{+-}_{\epsilon,\epsilon'}
(t^h_{\epsilon}-t^h_{\epsilon'})
\begin{pmatrix}
0 \\
\bar{\Sigma}^{a,+}_{\epsilon,\epsilon'(\omega_1,\omega_2)} 
\end{pmatrix}. 
\end{split}
\label{eq:gbara}
\end{equation}
Here, $\bar{\Sigma}$ indicates the vertex correction, 
which is calculated in the next subsection, 
and the diagonal element ($\bar{\Sigma}^{n,+}$) 
is shown to vanish. 
\begin{equation}
\hat{M}^{ab}_{\epsilon,\epsilon'}
:=
\frac{i\left[
\hat{\tau}_3
-X^{ab}_{\epsilon,\epsilon'}\hat{\tau}_0
-Y^{ab}_{\epsilon,\epsilon'}\hat{\tau}_1
\right] 
}{z_{\epsilon}^a +z_{\epsilon'}^b
+2i\alpha_p X^{ab}_{\epsilon,\epsilon'}}. 
\end{equation}
Here, 
$X^{ab}_{\epsilon,\epsilon'}:= 
(\epsilon^a_p{\epsilon'}^b_p
+\Delta^a_{\epsilon}{\Delta}^b_{\epsilon'})/
z^a_{\epsilon}z^b_{\epsilon'}$, 
$Y^{ab}_{\epsilon,\epsilon'}:= 
(\epsilon^a_p \Delta_{\epsilon'}^b
+\Delta^a_{\epsilon}\epsilon^{'b}_p)/
z^a_{\epsilon}z^b_{\epsilon'}$, 
$\epsilon^{\pm}_p:=\epsilon-\alpha_p g^{\pm}_{\epsilon}$, 
$\Delta^{\pm}_{\epsilon}:=\Delta+\alpha_p f^{\pm}_{\epsilon}$, 
and 
$z^{\pm}_{\epsilon}:=\sqrt{(\epsilon_p^{\pm})^2
-(\Delta^{\pm}_{\epsilon})^2}$. 
Quasiclassical Green functions in the equilibrium state
are written as 
$g_{\epsilon}^{\pm}=-i\epsilon_p^{\pm}/z^{\pm}_{\epsilon}$
and
$f_{\epsilon}^{\pm}=-i\Delta_{\epsilon}^{\pm}/z^{\pm}_{\epsilon}$.

By using the above results, the nonlinear current 
is written as 
\begin{equation}
\begin{split}
 J_{\omega}=&-(e^2D_{\alpha})^2
\frac{mk_F}{2\pi}
\int\frac{dw dw'}{(2\pi)^3}
A_{\omega-2w}A_{w+w'/2}A_{w-w'/2}
I^{(3)}_{\omega,w,w'}.
\end{split}
\label{eq:pcurrent}
\end{equation}
Here, 
\begin{equation}
\begin{split}
I^{(3)}_{\omega,w,w'}:=&
\int\frac{d\epsilon}{4\pi i}
{\rm Tr}
\Bigl[ 
\left(
\hat{g}^+_{\epsilon+\omega-w}
-\hat{\tau}_3\hat{g}^+_{\epsilon+w}\hat{\tau}_3
+\hat{g}^-_{\epsilon-\omega+w}
-\hat{\tau}_3\hat{g}^-_{\epsilon-w}\hat{\tau}_3\right)
\hat{\bar{g}}^{K}_{\epsilon+w,\epsilon-w(w+w'/2,w-w'/2)} \\
&+
\left(
\hat{g}^K_{\epsilon+\omega-w}
-\hat{\tau}_3\hat{g}^K_{\epsilon+w}\hat{\tau}_3
\right)
\hat{\bar{g}}^{-}_{\epsilon+w,\epsilon-w(w+w'/2,w-w'/2)}
+
\left(
\hat{g}^K_{\epsilon-\omega+w}
-\hat{\tau}_3\hat{g}^K_{\epsilon-w}\hat{\tau}_3
\right)
\hat{\bar{g}}^{+}_{\epsilon+w,\epsilon-w(w+w'/2,w-w'/2)}
\Bigr]. 
\end{split}
\label{eq:smalli3}
\end{equation}

\subsection{Vertex correction}

We obtain $\Sigma^{(ep)+}_{\epsilon,\epsilon'}$ 
by solving 
\begin{equation}
\begin{pmatrix}
\Sigma^{(ep)n,+}_{\epsilon,\epsilon'} \\
\Sigma^{(ep)a,+}_{\epsilon,\epsilon'} 
\end{pmatrix}
=
g_0\int\frac{d\epsilon_1}{2\pi i}
\hat{\tau}_3
\left[
\begin{pmatrix}
g^{(a)}_{\epsilon_1+w,\epsilon_1-w} \\
f^{(a)}_{\epsilon_1+w,\epsilon_1-w}
\end{pmatrix}
+
t^h_{\epsilon_1-w}
\begin{pmatrix}
g^{+}_{\epsilon_1+w,\epsilon_1-w} \\
f^{+}_{\epsilon_1+w,\epsilon_1-w}
\end{pmatrix}
-
t^h_{\epsilon_1+w}
\begin{pmatrix}
g^{-}_{\epsilon_1+w,\epsilon_1-w} \\
f^{-}_{\epsilon_1+w,\epsilon_1-w}
\end{pmatrix}
\right]
\end{equation}
[$w:=(\epsilon-\epsilon')/2$]. 
Using Eqs. (\ref{eq:GexclA})-(\ref{eq:gbara}) and 
\begin{equation}
\begin{pmatrix}
\Sigma^{(ep)n,+}_{\epsilon,\epsilon'} \\ 
\Sigma^{(ep)a,+}_{\epsilon,\epsilon'} 
\end{pmatrix}
:=e^2D_{\alpha}
\int\frac{d\omega_1d\omega_2}{(2\pi)^2}
A_{\omega_1}A_{\omega_2}\delta(\epsilon-\epsilon'-\omega_1-\omega_2)
\begin{pmatrix}
\bar{\Sigma}^{n,+}_{\epsilon,\epsilon'(\omega_1,\omega_2)} 
\\ \bar{\Sigma}^{a,+}_{\epsilon,\epsilon'(\omega_1,\omega_2)} 
\end{pmatrix}, 
\label{eq:SexclA}
\end{equation}
the solution is written as 
\begin{equation}
\begin{split}
\begin{pmatrix}
\bar{\Sigma}^{n,+}_{\epsilon,\epsilon'(\omega_1,\omega_2)} \\
\bar{\Sigma}^{a,+}_{\epsilon,\epsilon'(\omega_1,\omega_2)} 
\end{pmatrix}
=
\frac{g_0}{2D_{2w}}\int\frac{d\epsilon_1}{2\pi i}
\begin{pmatrix}
0 & 0 \\ 0 & -1
\end{pmatrix}
\sum_{s,s'=\pm}
s\Bigl[ &
(t^h_{\epsilon_1-s'w'}-t^h_{\epsilon_1-sw})
\hat{M}^{+-}_{\epsilon_1+w,\epsilon_1-w} \\
&+
t^h_{\epsilon_1-sw}
\hat{M}^{ss}_{\epsilon_1+w,\epsilon_1-w}
\Bigr]
\begin{pmatrix}
g^s_{\epsilon_1-s'w'} \\
f^s_{\epsilon_1-s'w'}
\end{pmatrix}
\end{split}
\label{eq:sfcorr}
\end{equation}
with $w=\epsilon-\epsilon'=(\omega_1+\omega_2)/2$ 
and $w'=(\omega_1-\omega_2)/2$. 
Here, 
\begin{equation}
D_{2w}=
1-g_0\int\frac{d\epsilon_1}{2\pi i}
\left[
(t^h_{\epsilon_1+w}-t^h_{\epsilon_1-w})
m^{+-}_{\epsilon_1+w,\epsilon_1-w}
+
t^h_{\epsilon_1-w}
m^{++}_{\epsilon_1+w,\epsilon_1-w}
-
t^h_{\epsilon_1+w}
m^{--}_{\epsilon_1+w,\epsilon_1-w}
\right]
\label{eq:ampmode}
\end{equation}
with 
\begin{equation}
m^{ab}_{\epsilon,\epsilon'}:=
\frac{i\left(1+X^{ab}_{\epsilon,\epsilon'}\right)
}{z_{\epsilon}^a +z_{\epsilon'}^b
+2i\alpha_pX^{ab}_{\epsilon,\epsilon'}}. 
\end{equation}

\section{Third-Harmonic Generation}

\subsection{Nonlinear response under incident pulse}

The electric field is written as 
$\tilde{E_t}=\bar{E_t}e^{-i\Omega t}
+\bar{E_t}^*e^{i\Omega t}$. 
Using of this field, 
the vector potential is written as 
$A_{\omega}=E_{\omega}/i\omega$ 
with 
$E_{\omega}=\int\frac{dt}{2\pi}\tilde{E_t}e^{i\omega t}$. 
We assume a Gaussian pulse for $\bar{E_t}$: 
$|\bar{E_t}|=\bar{E_0}e^{-(t/t_0)^2}$. 
Then
\[
A_{\omega}=\frac{\sqrt{\pi}t_0}{i\omega}
\left[
\bar{E_0}e^{-(\omega-\Omega)^2/4c}+
\bar{E_0}^*e^{-(\omega+\Omega)^2/4c}\right], 
\]
and we introduce a dimensionless external field, 
\[
\bar{A}_{\omega}:=
\frac{\Delta_0}{\sqrt{\pi}t_0|\bar{E_0}|}A_{\omega}. 
\]
($\Delta_0$ is the superconducting gap 
at $T=0$ without impurity scatterings.) 
Using of $\bar{A}_{\omega}$, 
the current [Eq. (\ref{eq:pcurrent})] is 
rewritten as 
\begin{equation}
J_{\omega}=\sigma_0 |E_{\Omega}|
\left(\frac{e|\bar{E_0}|\xi_0}{\Delta_0}\right)^2
\frac{l}{\xi_0}j_{\omega} 
\label{eq:currentJ}
\end{equation}
with 
\begin{equation}
 j_{\omega}:=\frac{-1}{48}
\int dw dw't_0^2
\bar{A}_{\omega-2w}\bar{A}_{w+w'/2}\bar{A}_{w-w'/2}
I^{(3)}_{\omega,w,w'}. 
\label{eq:smallj}
\end{equation}
Here,
$|E_{\Omega}|=\sqrt{\pi}t_0|\bar{E_0}+\bar{E_0}^*
e^{-(\Omega t_0)^2}|\simeq
\sqrt{\pi}t_0|\bar{E_0}|$ and 
$\sigma_0=n_e e^2 \tau/m=e^2 D_{\alpha} mk_F/\pi^2$ 
are used. 
$\xi_0=v_F/\pi\Delta_0$ and 
$l=v_F\tau$ 
are the coherence length and 
the mean free path, respectively. 

Equation (\ref{eq:smalli3}) does not 
depend on $\alpha$. 
This is because 
$\hat{g}^s_{\epsilon}$ 
and $\hat{\bar{g}}^s_{\epsilon,\epsilon'}$ 
do not include $\alpha$ owing to 
Anderson's theorem~\cite{anderson59} and 
the absence of a collision term derived from nonmagnetic impurity 
scattering~\cite{vollhardt,jujo17}, respectively. 
The dependence of the nonlinear current 
[Eq. (\ref{eq:currentJ})] on nonmagnetic impurity scattering 
is included only in $l$. 
Thus, the following numerical calculations are performed 
without specifying the value of $\alpha$.

\subsection{Numerical calculations}

We calculate the THG intensity using 
the dimensionless $j_{\omega}$ from Eq. (\ref{eq:smallj}). 
We take $\Delta_0$ as the unit of energy ($\Delta_0=1$), 
and in the variables in the subsequent figures, 
the notation of $\Delta_0$ 
is omitted for the sake of simplicity. 

\begin{figure}
\includegraphics[width=11.5cm]{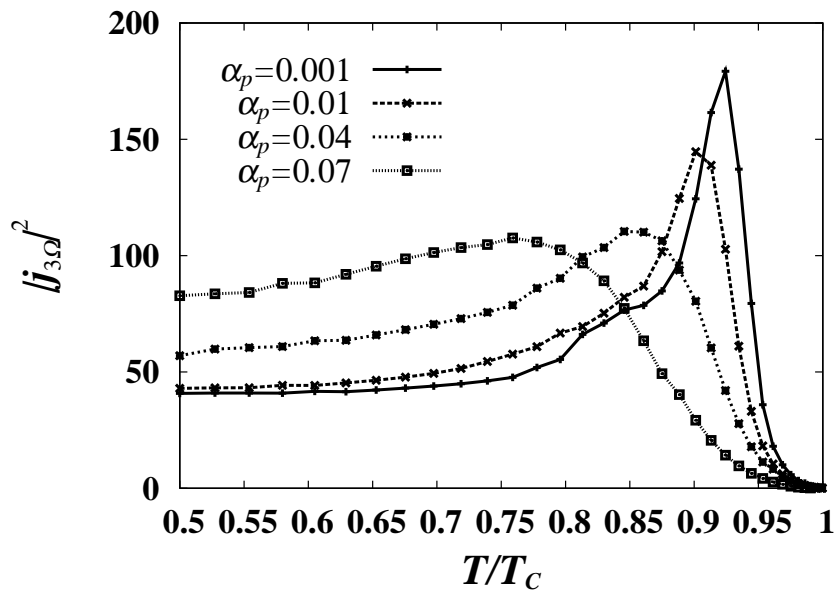}
\caption{\label{fig:1} 
Dependence of 
the dimensionless THG intensity $|j_{3\Omega}|^2$ 
on temperature ($T$ normalized by the transition 
temperature $T_C$) for various values of 
$\alpha_p$ (the effect of paramagnetic impurities). 
$\Omega=0.46\Delta_0$ (the frequency of the incident pulse). 
$1/t_0^2=0.003\Delta_0^2$. 
(This value corresponds to the full width 
at half maximum 
$\delta\Omega=4\sqrt{{\rm ln}2}/t_0
\simeq 0.18\Delta_0$.)}
\end{figure}

The dependence of the THG intensity ($|j_{3\Omega}|^2$ 
with $\Omega$ the frequency of the incident pulse) 
on temperature is shown in Fig.~\ref{fig:1}. 
There is a sharp peak for small values of $\alpha_p$. 
This peak becomes blurred and shifts to low temperatures 
with increasing $\alpha_p$. 
As shown below, this behavior is caused by the fact that 
this peak originates from the amplitude mode, 
and the blurring occurs owing to 
the vanishing of the well-defined mode for 
finite values of $\alpha_p$. 

\begin{figure}
\includegraphics[width=11.5cm]{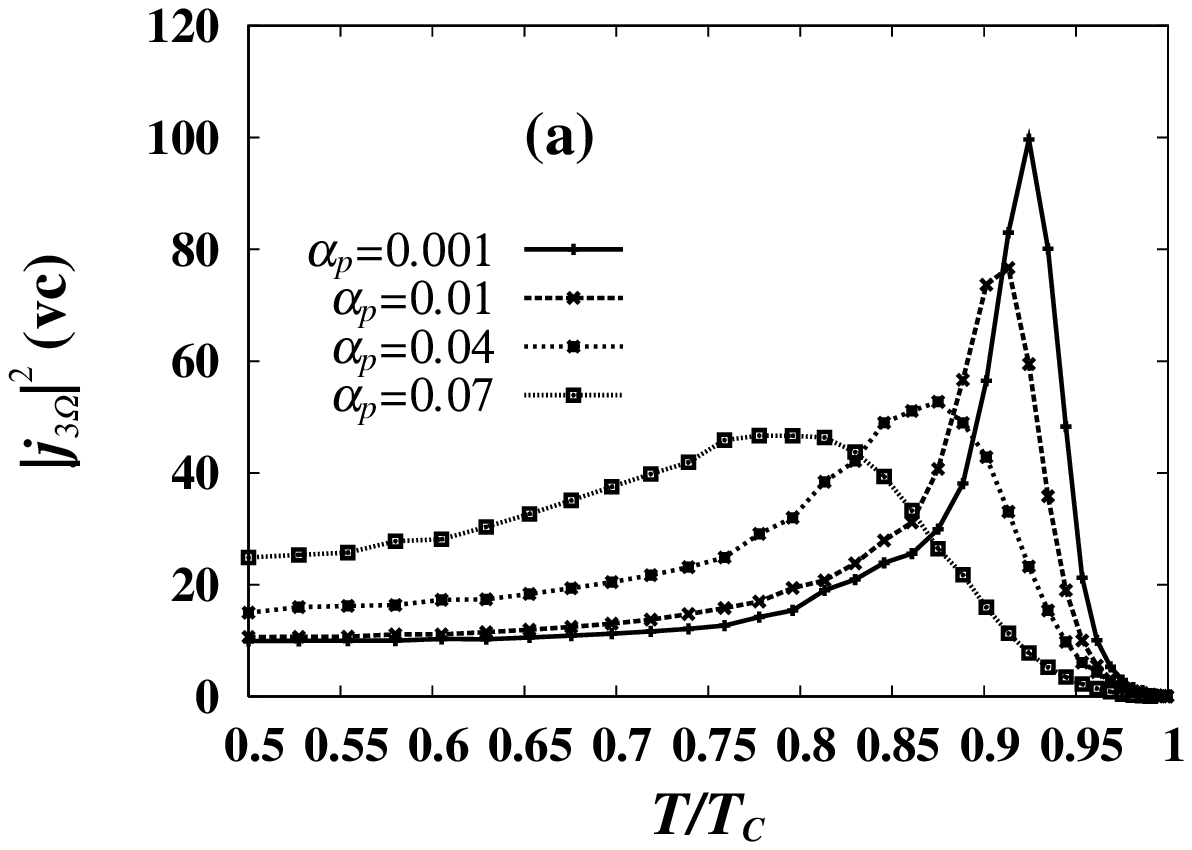}
\includegraphics[width=11.5cm]{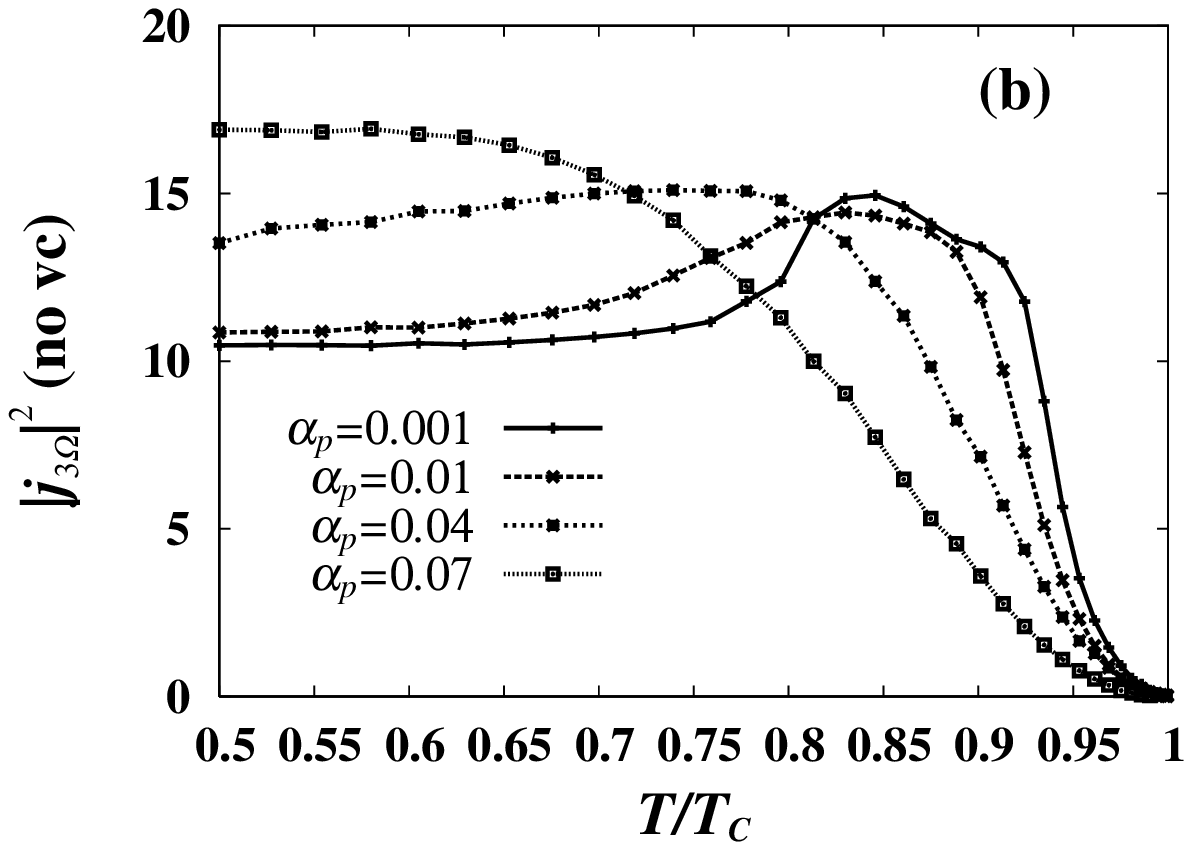}
\caption{\label{fig:2} (a) 
Temperature dependence of the THG intensity 
with only $\bar{\Sigma}^a$ included. 
(b) Temperature dependence of 
the THG intensity without $\bar{\Sigma}^a$. 
$\Omega=0.46\Delta_0$ and $1/t_0^2=0.003\Delta_0^2$.}
\end{figure}

The amplitude mode described by $D_{2\omega}$ 
[Eq. (\ref{eq:ampmode})] 
is included in the vertex correction 
$\bar{\Sigma}^a$ [Eq. (\ref{eq:sfcorr})]. 
The THG intensities with and 
without the vertex correction are shown 
in Figs.~\ref{fig:2}(a) and 2(b), respectively. 
Figure 2(a) [2(b)] is calculated using $\bar{g}_{\omega}$, 
in which only the second [first] terms 
in the left-hand side of Eqs. (\ref{eq:gbar+}) and (\ref{eq:gbara}) 
are included. 
The THG intensity is proportional to the square of 
the absolute value of $j_{3\Omega}$, and thus, 
the result in Fig. 1 is not a simple summation of 
the results in Figs. 2(a) and 2(b). 
Figure 2(a) indicates that 
the vertex correction causes a sharp peak 
for small values of $\alpha_p$. 

\begin{figure}
\includegraphics[width=11.5cm]{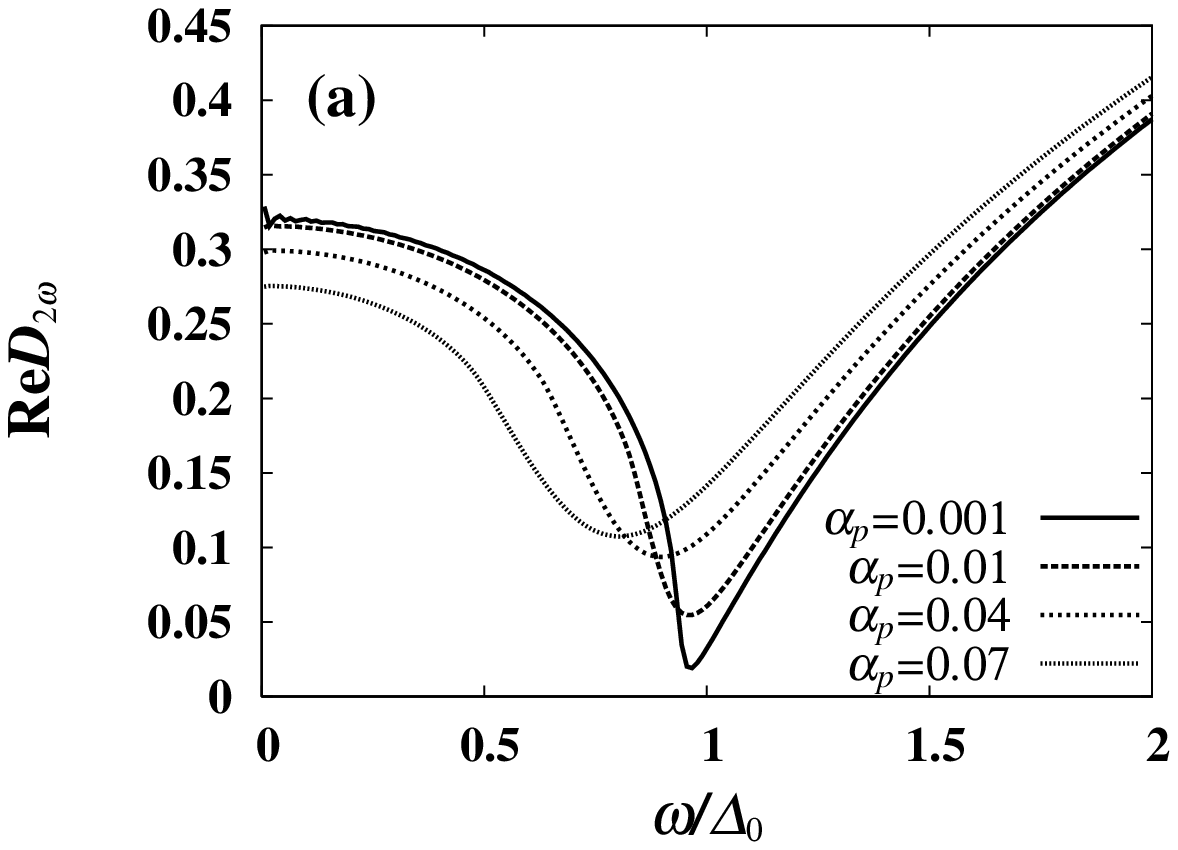}
\includegraphics[width=11.5cm]{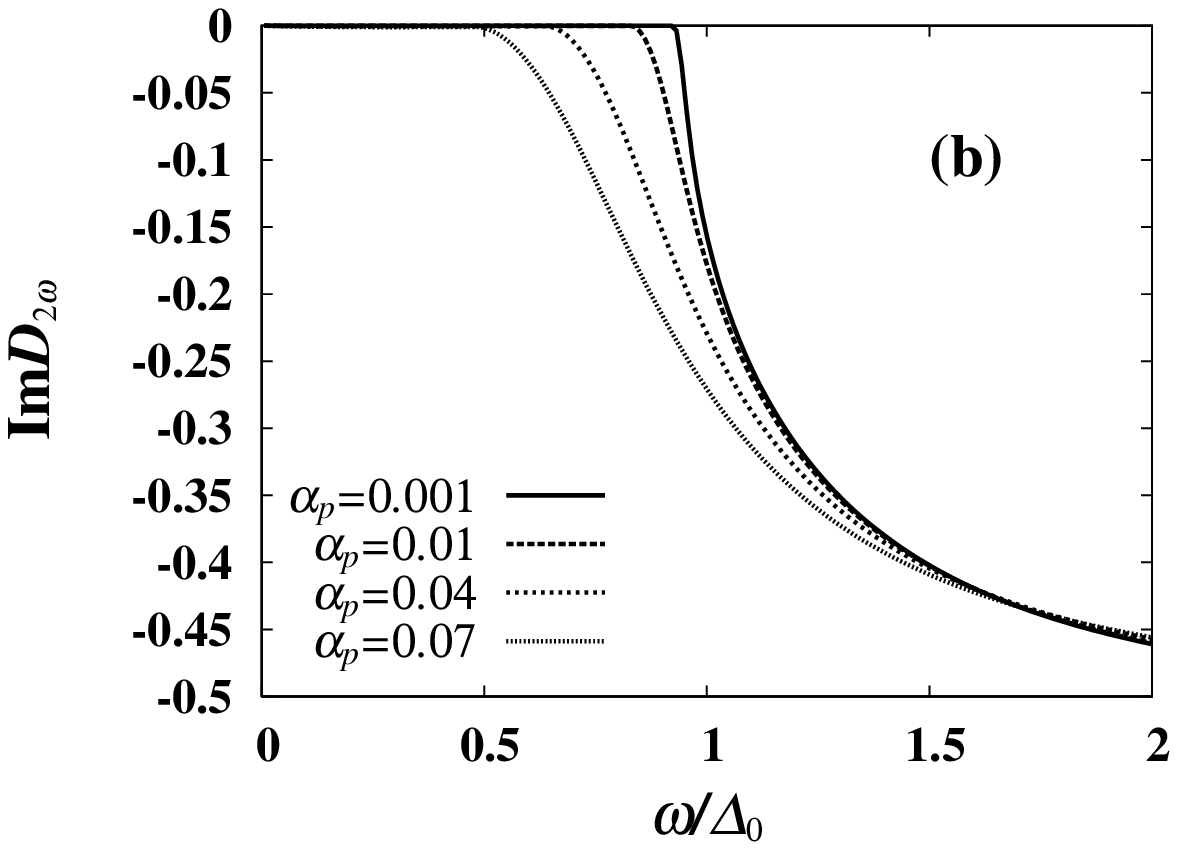}
\caption{\label{fig:3}(a) Frequency ($\omega$) dependence of 
the real part of $D_{2\omega}$ for various values of 
$\alpha_p$. 
(b) Frequency dependence of 
the imaginary part of $D_{2\omega}$. 
$D_{2\omega}$ is a dimensionless quantity. 
$T/T_C=0.5$.} 
\end{figure}

The dependence of the amplitude mode ($D_{2\omega}$) 
on the frequency $\omega$ at $T/T_C=0.5$ is shown in Fig.~\ref{fig:3}.
Figure 3(a) shows that a well-defined amplitude mode 
disappears with increasing $\alpha_p$. 
$-{\rm Im}D_{2\omega}$ is the damping rate of 
the amplitude mode and takes finite values 
in the range $\omega<\Delta$ for finite values of $\alpha_p$. 
This behavior originates from the fact that 
the gap edge ($E_g$) of a one-particle spectrum 
is smaller than $\Delta$ for finite values of $\alpha_p$: 
$E_g=\Delta[1-(2\alpha_p/\Delta)^{2/3}]^{3/2}$ 
(see Ref. 19). 
%~\cite{skalski}). 

\begin{figure}
\includegraphics[width=9.5cm]{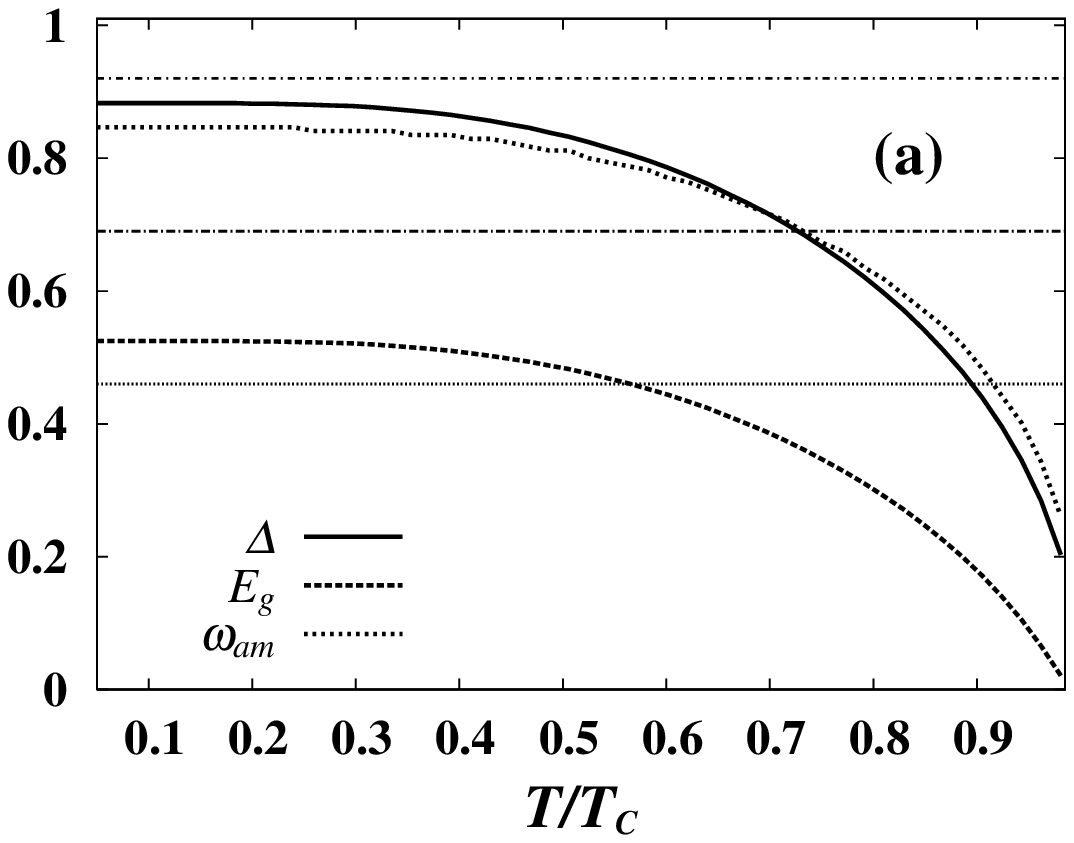}
\includegraphics[width=9.5cm]{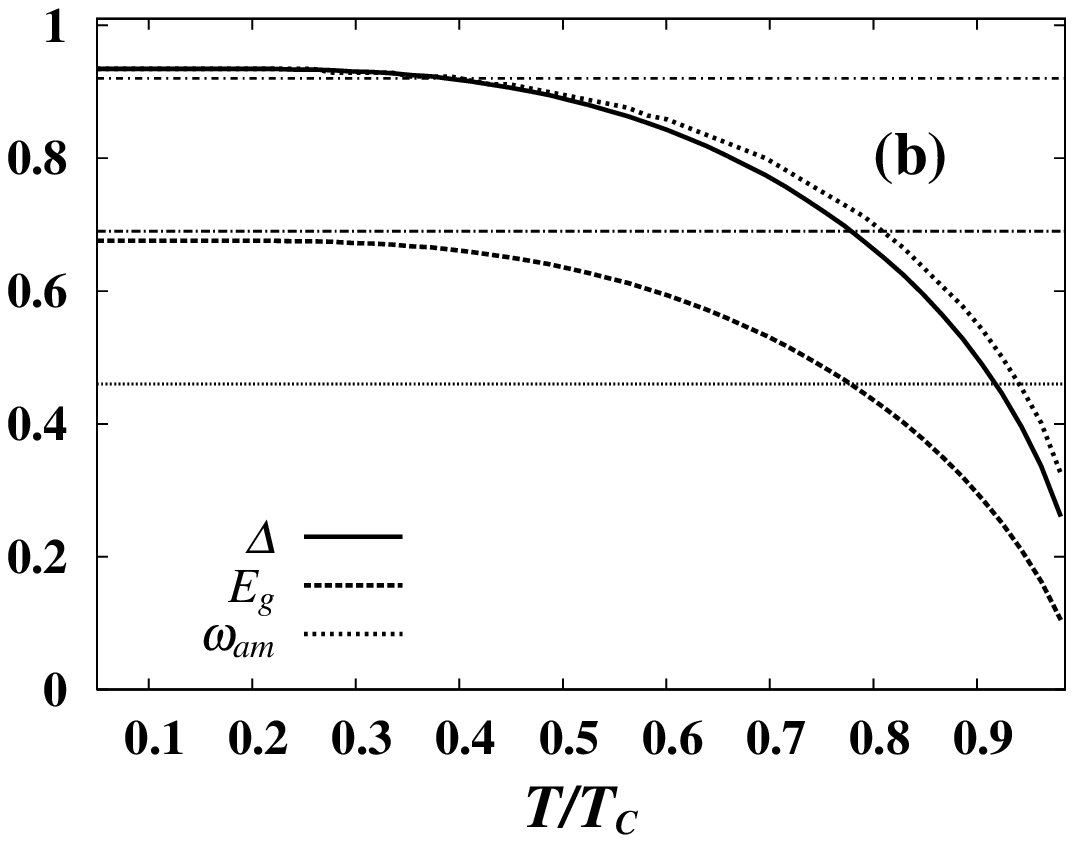}
\includegraphics[width=9.5cm]{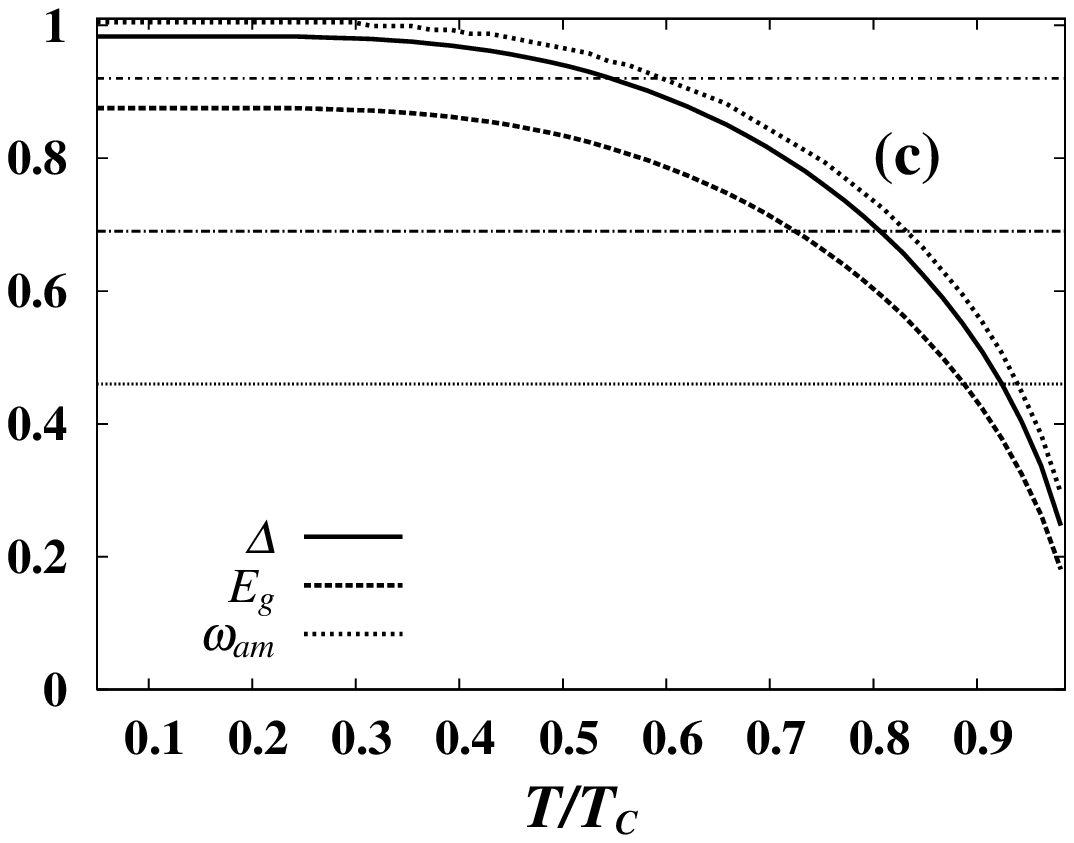}
\includegraphics[width=9.5cm]{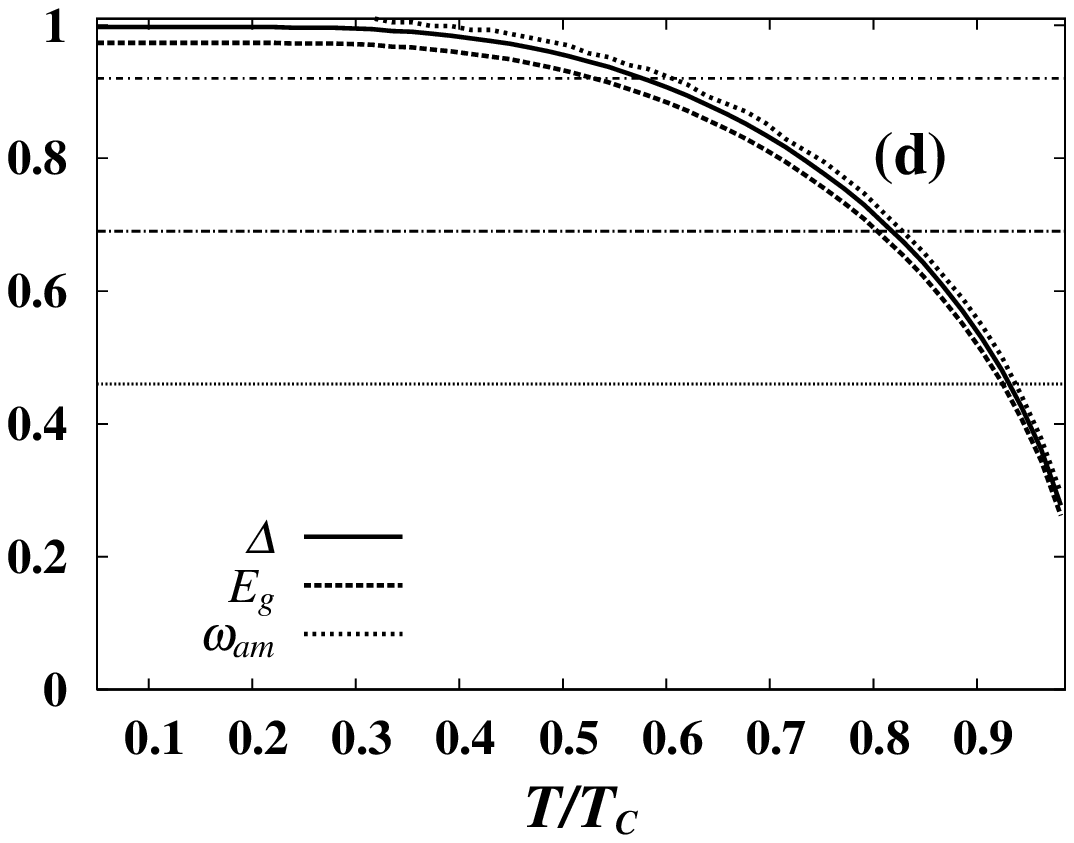}
\caption{\label{fig:4} Temperature dependences of 
$\Delta$, $E_g$, and the frequency 
$\omega_{am}$ (at which ${\rm Re}D_{2\omega}$ takes the local 
minimum). 
(a) $\alpha_p=0.07\Delta_0$. 
(b) $\alpha_p=0.04\Delta_0$. (c) $\alpha_p=0.01\Delta_0$. 
(d) $\alpha_p=0.001\Delta_0$. 
The horizontal lines indicate the values 
($\Omega=$) $0.46\Delta_0$, $0.69\Delta_0$, and $0.92\Delta_0$ 
from bottom to top.} 
\end{figure}

We introduce $\omega_{am}$ as the frequency 
at which ${\rm Re}D_{2\omega}$ takes a local minimum. 
The dependence of $\omega_{am}$ on temperature 
is shown in Fig.~\ref{fig:4}. 
The values of $\omega_{am}$ are numerically calculated 
from the dependences of ${\rm Re}D_{2\omega}$ on frequency. 
Figure 4 shows that 
$\omega_{am}\simeq \Delta$ regardless of the value of $\alpha_p$. 
In contrast, 
$E_g$ deviates from $\Delta$ with increasing $\alpha_p$ as noted 
above. 
The reason for including horizontal lines in Fig. 4 is 
to investigate how the peak of the THG intensity is related 
to the temperature at which the frequency ($\Omega$) crosses $\Delta$ 
or $E_g$. 

\begin{figure}
\includegraphics[width=11.5cm]{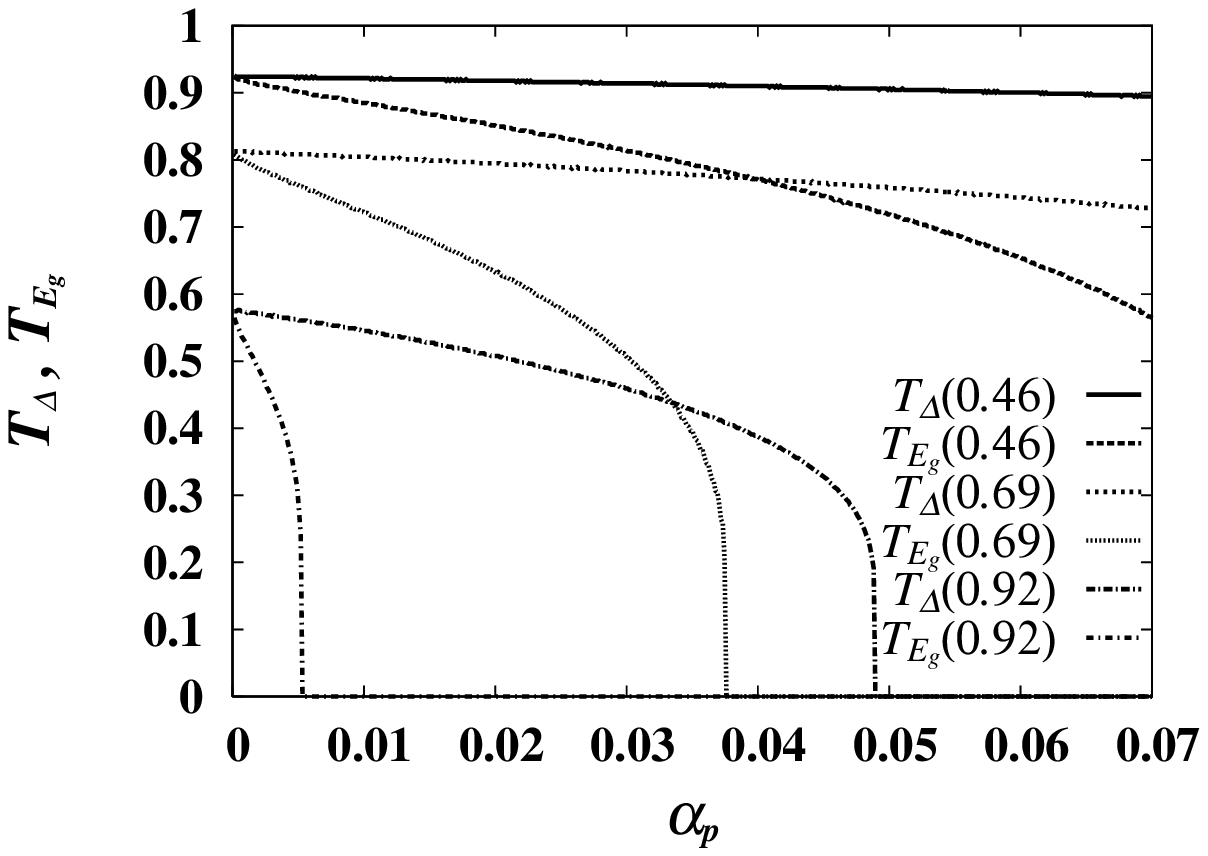}
\caption{\label{fig:5} $T_{\Delta}$ ($T_{E_g}$), 
which is the intersection of 
$\Delta$ ($E_g$) and $\Omega$. 
Numbers in parentheses are values of $\Omega$. 
$T_{\Delta}$ and $T_{E_g}$ are values of the temperature normalized by $T_C$.} 
\end{figure}

The temperature (normalized by $T_C$) 
at which $\Delta$ (or $E_g$) and $\Omega$ ($=0.46\Delta_0$, $0.69\Delta_0$, 
$0.92\Delta_0$) intersect is shown in Fig.~\ref{fig:5}. 
($T_{\Delta}$ and $T_{E_g}$ are dimensionless quantities.) 
$T_{\Delta}$ and $T_{E_g}$ are set to 0 
when $\Delta$ and $E_g$ do not 
cross $\Omega$, respectively. 
Figures 1 and 5 show that 
the THG intensity does not show a peak 
at $T/T_C=T_{\Delta}$ or $T/T_C=T_{E_g}$. 
The temperature at which the THG intensity shows a peak 
is between $T_{\Delta}$ and $T_{E_g}$. 

\begin{figure}
\includegraphics[width=11.5cm]{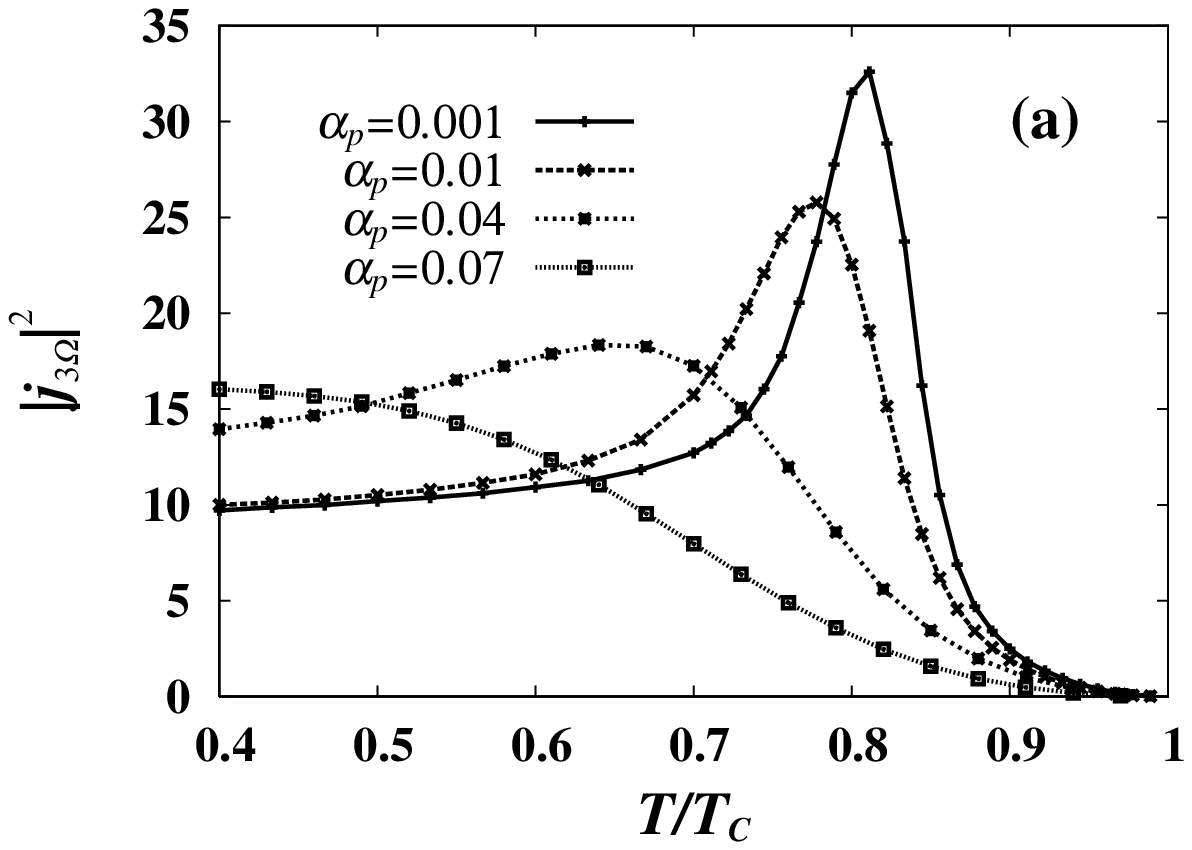}
\includegraphics[width=11.5cm]{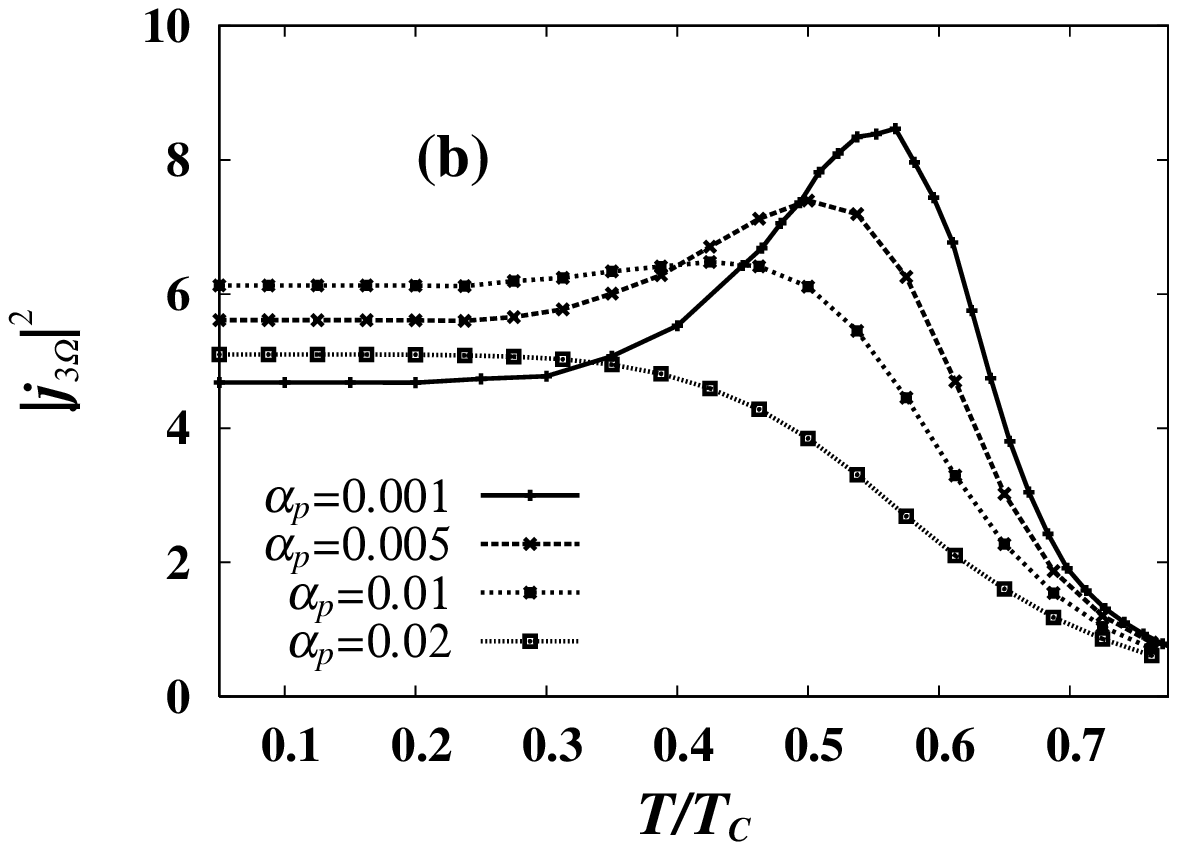}
\caption{\label{fig:6} 
Temperature dependences of the dimensionless 
THG intensity $|j_{3\Omega}|^2$ for various values of 
$\alpha_p$ with $\Omega$ the frequency of the incident pulse. 
(a) $\Omega=0.69\Delta_0$ and (b) $\Omega=0.92\Delta_0$. 
$1/t_0^2=0.003\Delta_0^2$.}
\end{figure}

This is verified by calculating the 
THG intensity for other values of $\Omega$. 
The THG intensities for $\Omega=0.69\Delta_0$ 
and $\Omega=0.92\Delta_0$ are shown 
in Figs.~\ref{fig:6}(a) and 6(b), respectively. 
Figures 6(a) and 6(b) show that the peak of the THG intensity 
almost vanishes in the case of $T_{E_g}=0$. 
The deviation of the peak from $T_{\Delta}$ 
is similar to the case of the two-photon absorption (TPA) spectrum. 
In the latter case, 
the frequency at which 
the TPA spectrum shows a peak 
is between $E_g$ and $\Delta$.~\cite{jujolt28} 
This behavior is caused by a shift of the spectrum 
to a lower energy with increasing $\alpha_p$. 

\begin{figure}
\includegraphics[width=11.5cm]{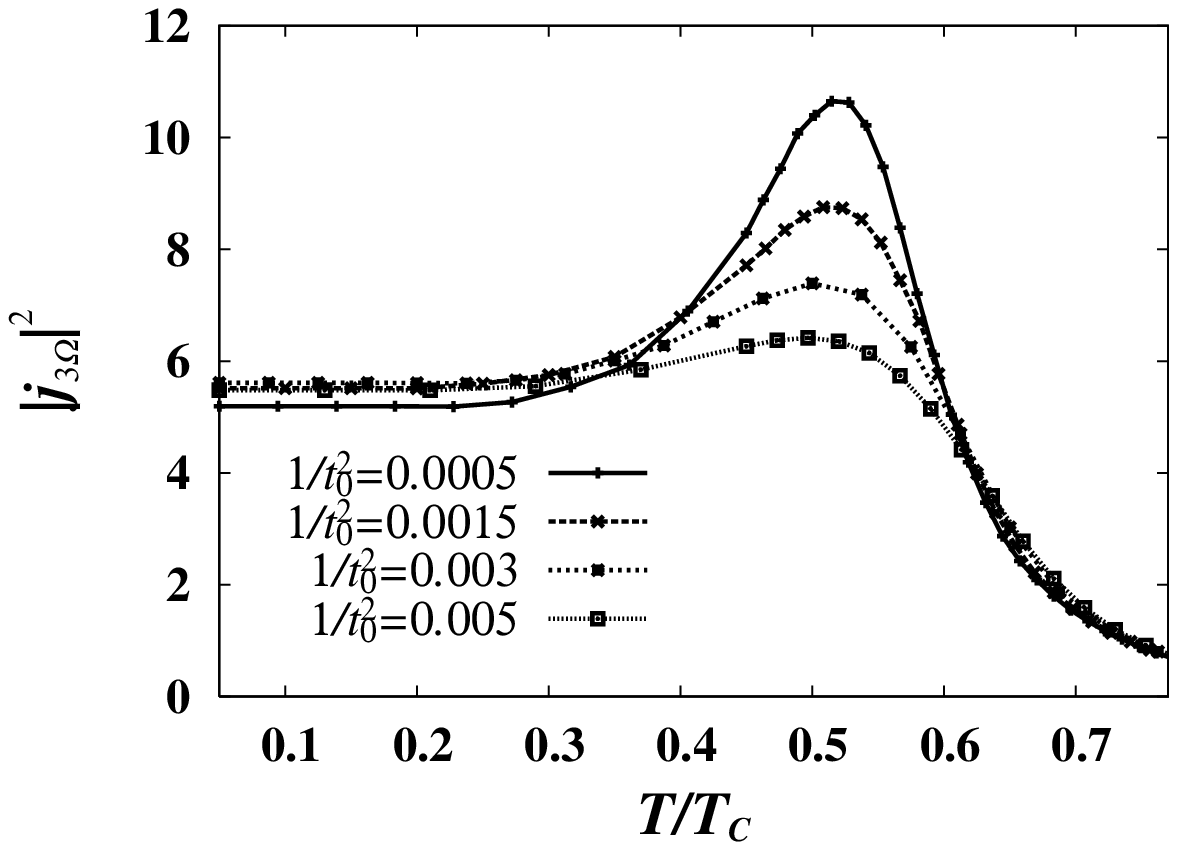}
\caption{\label{fig:7} 
Dependence of the dimensionless THG intensity 
on temperature for various values of 
$1/t_0^2$ (with $1/t_0$ proportional to 
the pulse width in the frequency space). 
$\alpha_p=0.005\Delta_0$ and $\Omega=0.92\Delta_0$.} 
\end{figure}

The THG intensities for various values of $1/t_0^2$ are shown 
in Fig.~\ref{fig:7}. 
A small value of $1/t_0^2$ corresponds to 
a small width of the incident pulse in the frequency space. 
The temperature at which the THG intensity shows a peak 
does not vary with $1/t_0^2$, 
but the peak becomes sharp with decreasing $1/t_0^2$. 
This behavior supports an expectation about the peak 
in the THG intensity suggested in Ref. 8. 
%~\cite{matsunaga14}. 

\subsection{Quantitative evaluation}

The THG intensity divided by the incident pulse 
is approximately written as 
\[
\left|
\frac{4\pi d}{2n_{3\Omega} c}\right|^2
\frac{|J_{3\Omega}|^2}{|E_{\Omega}|^2}
=
\left|\frac{4\pi d\sigma_0}{2n_{3\Omega} c}\right|^2
\left(\frac{e|\bar{E_0}|\xi_0}{\Delta_0}\right)^4
\left(\frac{l}{\xi_0}\right)^2
|j_{3\Omega}|^2 
\]
(see, for example, Ref. 24). 
%~\cite{shen}). 
Here, $n_{3\Omega}$ is the refractive index, 
$|n_{3\Omega}|^2
\simeq 4\pi |\sigma_{3\Omega}|/3\Omega$, and 
$d$ is the thickness of the film. 
When we set $\xi_0=5$ nm,~\cite{ikebe} $\Delta_0=2.7$ meV, 
$d=24$ nm, and $\sigma_0=1.5\times 10^4$ $\Omega^{-1}$ cm$^{-1}$, 
and assume $|\sigma_{3\Omega}|\simeq \sigma_0$ because we consider 
the case of $3\Omega>\Delta$, we obtain 
\[
 \left|\frac{4\pi d\sigma_0}{2n_{3\Omega} c}\right|^2
\left(\frac{e|\bar{E_0}|\xi_0}{\Delta_0}\right)^4
\left(\frac{l}{\xi_0}\right)^2
\simeq
1.37\times 10^{-5} 
\]
for 
$l=0.8$ nm ($l\simeq 0.58 \sim 0.83$ nm~\cite{semenov}), 
$|\bar{E_0}|=3.5$ kV/cm, and $\Omega=0.92\Delta_0$. 
Then 
$\left|\frac{4\pi d\sigma_0}{2n_{3\Omega} c}\right|^2
\left(\frac{e|\bar{E_0}|\xi_0}{\Delta_0}\right)^4
\left(\frac{l}{\xi_0}\right)^2
|j_{3\Omega}|^2 \simeq 6.85\times 10^{-5}$ 
using the calculated results of $|j_{3\Omega}|^2$ 
in Fig. 6(b) for $\Omega=0.92\Delta_0$. 
This is about the same value as an experimental value in 
Ref. 8, 
%~\cite{matsunaga14} 
where $8\times 10^{-5}$ was reported as 
a value of the THG intensity normalized by that of 
the pump pulse.

\section{Diamagnetic Term}

When we take account of the diamagnetic coupling 
in the interaction between electrons and 
external fields~\cite{note1}, there is an additional term 
in a nonlinear current as follows: 
\begin{equation}
 J^d_{\omega}=\frac{-e^2}{m}\int\frac{d\omega'}{2\pi}
A_{\omega'}\int_{\rm FS}\frac{mk_F}{2\pi}\int\frac{d\epsilon}{4\pi i}
{\rm Tr}\left[
\hat{\tau}_3\hat{g}^{K}_{\epsilon+\omega-\omega',\epsilon}
\right]. 
\label{eq:nlcJd}
\end{equation}
In this case, 
there are also additional terms in the kinetic equations. 
These are written as 
\begin{equation}
\int\frac{d\omega_1d\omega_2}{(2\pi)^2}
\frac{e^2}{2}
A_{\omega_1}A_{\omega_2}
\frac{1}{m}
\left[
\hat{g}^+_{\epsilon}-\hat{g}^+_{\epsilon'}
\right]\delta(\epsilon-\epsilon'-\omega_1-\omega_2)
\end{equation}
and 
\begin{equation}
\int\frac{d\omega_1d\omega_2}{(2\pi)^2}
\frac{e^2}{2}
A_{\omega_1}A_{\omega_2}
\frac{1}{m}(t^h_{\epsilon}-t^h_{\epsilon'})
\left[
\hat{g}^+_{\epsilon}-\hat{g}^-_{\epsilon'}
\right]\delta(\epsilon-\epsilon'-\omega_1-\omega_2)
\end{equation}
for the left-hand sides of 
Eqs. (\ref{eq:kefor+}) and (\ref{eq:kefora}), 
respectively. 
The additional term arises in the solution 
[Eq. (\ref{eq:solg})] 
and is written as~\cite{note2}  
\begin{equation}
g'_{\epsilon,\epsilon'}\hat{\tau}_3
+if'_{\epsilon,\epsilon'}\hat{\tau}_2
\label{eq:solgd}
\end{equation} 
with $i\hat{\tau}_2=\left(
\begin{smallmatrix} 0 & 1 \\ -1 & 0 \end{smallmatrix}
\right)$. 

Since $(g',f')$ are different matrix elements from 
$(g,f)$ in Sects. 2 and 3, 
they can be obtained separately. 
The solution is written as 
\begin{equation}
\begin{pmatrix}
g^{'s}_{\epsilon,\epsilon'} \\ f^{'s}_{\epsilon,\epsilon'} 
\end{pmatrix}
=
\int\frac{d\omega_1d\omega_2}{(2\pi)^2}
\frac{e^2}{2}
A_{\omega_1}A_{\omega_2}\delta(\epsilon-\epsilon'-\omega_1-\omega_2)
\frac{1}{m}
\begin{pmatrix}
\bar{g}^{'s}_{\epsilon,\epsilon'} \\ \bar{f}^{'s}_{\epsilon,\epsilon'} 
\end{pmatrix}
\label{eq:bargp}
\end{equation}
with 
\begin{equation}
\begin{pmatrix}
\bar{g}^{'\pm}_{\epsilon,\epsilon'} \\ \bar{f}^{'\pm}_{\epsilon,\epsilon'} 
\end{pmatrix}
:=
\hat{M}^{'\pm\pm}_{\epsilon,\epsilon'}
\begin{pmatrix}
1\\0
\end{pmatrix} 
\end{equation}
and 
\begin{equation}
\begin{pmatrix}
\bar{g}^{'(a)}_{\epsilon,\epsilon'} \\ \bar{f}^{'(a)}_{\epsilon,\epsilon'} 
\end{pmatrix}
:=
(t^h_{\epsilon}-t^h_{\epsilon'})
\hat{M}^{'+-}_{\epsilon,\epsilon'}
\begin{pmatrix}
1\\0
\end{pmatrix}. 
\end{equation}
Here, 
\begin{equation}
\hat{M}^{'ab}_{\epsilon,\epsilon'}:=
\frac{i\left[
\hat{\tau}_3
-X^{'ab}_{\epsilon,\epsilon'}\hat{\tau}_0
-Y^{'ab}_{\epsilon,\epsilon'}\hat{\tau}_1
\right] 
}{z_{\epsilon}^a +z_{\epsilon'}^b
+2i\alpha_p X^{'ab}_{\epsilon,\epsilon'}}, 
\end{equation}
where 
$X^{'ab}_{\epsilon,\epsilon'}:= 
(\epsilon^a_p{\epsilon'}^b_p
-\Delta^a_{\epsilon}{\Delta}^b_{\epsilon'})
/z^a_{\epsilon}z^b_{\epsilon'}$ 
and 
$Y^{'ab}_{\epsilon,\epsilon'}:= 
(\epsilon^a_p \Delta_{\epsilon'}^b
-\Delta^a_{\epsilon}\epsilon^{'b}_p)
/z^a_{\epsilon}z^b_{\epsilon'}$. 
Using the above quantities, 
Eq. (\ref{eq:nlcJd}) is rewritten as 
\begin{equation}
 J^{d}_{\omega}=
\frac{-1}{2}
\left(\frac{e^2}{m}\right)^2
\int\frac{dw dw'}{(2\pi)^3}
A_{\omega-2w}A_{w+w'/2}A_{w-w'/2}
\frac{mk_F}{2\pi}\int\frac{d\epsilon}{4\pi i}
{\rm Tr}
\Bigl[\hat{\tau}_3
\hat{\bar{g}}^{'K}_{\epsilon+w,\epsilon-w}
\Bigr]. 
\end{equation}

By using the electric field introduced in Sect. 3, 
the nonlinear current is written as 
\begin{equation}
 J^{d}_{\omega}=
\sigma_0|E_{\Omega}|
\left(\frac{e|\bar{E_0}|\xi_0}{\Delta_0}\right)^2
\left(\frac{\Delta_0}{E_F}\right)^2
\frac{9\pi^2\xi_0}{8l}
j^{d}_{\omega}. 
\label{eq:currentJd}
\end{equation}
Here, $E_F=k_F^2/2m$ is the Fermi energy and 
\begin{equation}
 j^{d}_{\omega}=
\frac{-1}{48}
\int dw dw' t_0^2
\bar{A}_{\omega-2w}\bar{A}_{w+w'/2}\bar{A}_{w-w'/2}
\int\frac{d\epsilon}{4\pi i}
{\rm Tr}
\Bigl[\hat{\tau}_3
\hat{\bar{g}}^{'K}_{\epsilon+w,\epsilon-w}
\Bigr]. 
\end{equation}

We consider the ratio $|J_{3\Omega}^d|^2/|J_{3\Omega}|^2$
in order to evaluate this term quantitatively. 
From Eqs. (\ref{eq:currentJ}) and (\ref{eq:currentJd}), 
\begin{equation}
\frac{|J_{3\Omega}^d|^2}{|J_{3\Omega}|^2}
=\left(\frac{\Delta_0}{E_F}\right)^4
\left(\frac{3\pi \xi_0}{2\sqrt{2}l}\right)^4
\frac{|j_{3\Omega}^d|^2}{|j_{3\Omega}|^2}. 
\end{equation}
Here, 
$3\pi \xi_0/2\sqrt{2}l\simeq 20.8$ 
for $\xi_0=5$ nm and $l=0.8$ nm, and 
then 
$|J_{3\Omega}^d|^2/|J_{3\Omega}|^2\simeq 2\times 10^{-3}$
when $\Delta_0/E_F=0.01$ 
and $|j_{3\Omega}^d|^2/|j_{3\Omega}|^2=1$ 
(this last value is an overestimated value as mentioned below). 
Thus, the diamagnetic term is negligible 
as suggested in the annotation of Ref. 10 
%~\cite{jujo15} 
(Ref. 14 therein). 
This result is consistent with the experimental result 
that the diamagnetic term is not observed.~\cite{matsunaga17} 

The result of a numerical calculation shows that 
$|j_{3\Omega}^d|^2/|j_{3\Omega}|^2$ is smaller than 1. 
The reason for this is that 
there is no increase due to the amplitude mode in 
$|j^d_{3\Omega}|^2$. 
The absence of the amplitude mode in $|j^d_{3\Omega}|^2$ 
is the result of the properties of 
the diamagnetic coupling term, which is represented by 
$\hat{\tau}_3$ and $i\hat{\tau}_2$ 
in Eq. (\ref{eq:solgd}). 
($\hat{\tau}_3$ and $i\hat{\tau}_2$ indicate 
the density fluctuation and the phase mode, respectively. 
The amplitude mode in Sect. 2 comes from $\hat{\tau}_1$.)

\section{Summary and Discussion}

In this paper a theoretical study of the third-harmonic generation 
in dirty BCS superconductors was carried out. 
We calculated the temperature dependence and frequency dependence 
of the THG intensity, and showed that 
the vertex correction term including the amplitude mode 
is dominant. 
We introduced the effect of paramagnetic impurities, 
and showed that time-reversal symmetry breaking 
destabilizes the amplitude mode and that this effect is 
reflected in the THG intensity. 

We showed that the dependences of the THG intensity on temperature 
and pulse width reproduce experimental results 
at the limit where the effect of paramagnetic impurities 
is small. 
Quantitatively, almost the same result as the experimental 
result was obtained. 
In addition, it is known that the diamagnetic term is 
small in the experiment.~\cite{matsunaga17} 
The calculated result of Sect. 4 gives an explanation 
of this. 
This result is based on the fact that the reduction 
effect due to the superconducting gap being smaller 
than the Fermi energy is more dominant than the increase caused 
by the mean free path being shorter than the coherence length.

It is known that the effect of time-reversal symmetry breaking 
(due to paramagnetic impurities in this paper) 
can also be obtained by applying an in-plane magnetic field.~\cite{maki} 
Referring to Figs. 2(a) and 2(b), 
the dependence of the THG intensity on $\alpha_p$ differs between 
the term including the amplitude mode and that not including 
this mode. 
Therefore, the results of our calculation show that 
the presence of the amplitude mode can be 
confirmed experimentally by applying 
a magnetic field in the plane.

\section*{Acknowledgement}

The numerical computation in this work was carried out 
at the Yukawa Institute Computer Facility.


\begin{thebibliography}{9}
\bibitem{note0} 
In Ref. 8, 
%~\cite{matsunaga14} 
a theoretical work~\cite{tsuji} is cited as an explanation 
of the amplitude mode in the THG intensity. 
In this theory, however, the paramagnetic coupling 
term~\cite{note1} is not included because 
this calculation takes the clean and local limit. 
The linear absorption spectrum in the experiment~\cite{matsunaga14} 
indicates that this coupling term is effective 
because there is a clear gap structure in the spectrum. 
Thus, the paramagnetic coupling term cannot be neglected 
in calculations of response functions,~\cite{jujo15} and 
it makes a predominant contribution to the THG intensity 
as shown in Sect. 3.
\bibitem{note1} Terms such as 
the paramagnetic coupling and the diamagnetic coupling 
between electrons and electromagnetic fields 
are the same as those in Ref. 31.
\bibitem{note2} It is known that 
only the anisotropic term remains 
in the diamagnetic term.~\cite{abrikosov74} 
The value of the anisotropic term 
with vertex corrections included 
is at most several times the value of the isotropic term without vertex 
corrections. 
Thus, it is sufficient to evaluate the latter case 
in order to find out whether the diamagnetic term 
or the paramagnetic term dominates.
\end{thebibliography}
\end{document}